\documentclass[letterpaper,twocolumn,10pt]{article}
\usepackage{usenix,epsfig,bytefield}
\usepackage{algorithm,multicol,refcount}
\usepackage[noend]{algpseudocode}
\usepackage{ifthen}
\usepackage{xcolor}
\usepackage{booktabs}
\usepackage{color}
\usepackage{colortbl}
\usepackage{float}                           
\usepackage{bigdelim}
\usepackage{amsmath}
\usepackage{amsfonts}
\usepackage{amssymb}
\usepackage{caption}
\usepackage{subcaption}
\usepackage{lipsum}

\usepackage[hyphens]{url}
\usepackage[colorlinks,citecolor=blue,breaklinks=true]{hyperref}%
\usepackage{svg}

\begin{document}

\newcommand\hmm[1]{\ifnum\ifhmode\spacefactor\else2000\fi>1000 \uppercase{#1}\else#1\fi}
\renewcommand{\algorithmicrequire}{\textbf{Input:}}
\renewcommand{\algorithmicensure}{\textbf{Output:}}

\newcommand{\Problem}{\hmm{t}oken coloring}
\newcommand{\redeem}{\hmm{d}efloat}


\date{}

\title{\Large \bf ColorFloat: Constant space token coloring}

{\author{\normalfont{Ryan Zarick\footnotemark[1]{} \hspace{1em} Bryan Pellegrino \hspace{1em} Isaac Zhang\footnotemark[1]{} \hspace{1em} Thomas Kim \hspace{1em} Caleb Banister}
\vspace{0.4em} \\
LayerZero Labs Ltd.}
\maketitle

\footnotetext{~Inventors at LayerZero Labs Ltd.}
\let\thefootnote\relax\footnotetext{Copyright \copyright{~2023 LayerZero Labs Ltd. All rights reserved.}}

\begin{abstract}
We present ColorFloat, a family of O(1) space complexity algorithms that solve the problem of attributing (coloring) fungible tokens to the entity that minted them (\emph{minter}).
Tagging fungible tokens with metadata is not a new problem and was first formalized in the Colored Coins protocol.
In certain contexts, practical solutions to this challenge have been implemented and deployed such as NFT.
We define the \emph{fungible token coloring problem}, one specific aspect of the Colored Coins problem, to be the problem of retaining fungible characteristics of the underlying token while accurately tracking the attribution of fungible tokens to their respective minters.
Fungible token coloring has a wide range of Web3 applications. One application which we highlight in this paper is the onchain yield-sharing collateral-based stablecoin.
\end{abstract}
\section{Introduction}
\label{sec:introduction}

The Colored Coins protocol~\cite{coloredcoins, rosenfeld2012overview} is a well-known protocol to ``mark'' tokens with metadata attributes.
In this paper, we focus on solving one particular aspect of the Colored Coins problem that, to our knowledge, has no existing practical onchain solution.
We define the \emph{fungible token coloring problem} as the problem of attributing (coloring) a fungible token to the \emph{minter} that minted it and tracking this color as the token is transferred to different \emph{users'} wallets.
The Colored Coins protocol provides a theoretical solution to this problem by losslessly tracking, per-wallet, a mapping of token quantities to colors.
However, this hypothetical implementation would require O(N) storage \emph{per-wallet} for a system of N minters, and O(N) computation to iterate over this storage per transaction.
Thus, the Colored Coins protocol cannot solve this problem in a way that scales to many minters.
In this paper, we present a class of scalable, storage-efficient algorithms to attribute fungible tokens to minters: \emph{ColorFloat}.

We first formally define the entities involved in the fungible token coloring problem.
The \emph{minter} generates demand for \emph{users} to purchase fungible tokens while the per-chain \emph{token contract} facilitates the minting, burning and/or transacting of these tokens within a given blockchain.
Each minter is assigned a unique \emph{color} (e.g., a unique numeric ID), which identifies it within the chain.
In turn, any tokens attributed to this minter are associated with this color.
The token contract tracks the \emph{mint} of each minter, or the number of tokens that are attributed to that minter within a given chain.
Additionally, the \emph{vault contract} aggregates the mint across all domains (i.e., all blockchains) which we term the \emph{circulation}.
The mint of a specific color $c$ within the set of all valid colors $\mathbf{C}$ is notated $mint_c$, and the circulation of $c$ is notated $circulation_c$.
We define the \emph{attribution} of a color $c$ as the proportion of $circulation_c$ to the total circulation:

$$attribution_c = \frac{circulation_c}{\sum_{x \in C}{circulation_x}}$$

The goal of ColorFloat is to fairly and accurately track $attribution_c$ for all colors $c$ in the system.
Within this setting, the challenges are threefold: (1) tracking the global quantity of tokens of each color, (2) propagating this information as tokens are transferred between wallets, and (3) facilitating redemption of colored tokens.

\begin{figure}
    \centering
    \includegraphics[width=1.00\columnwidth]{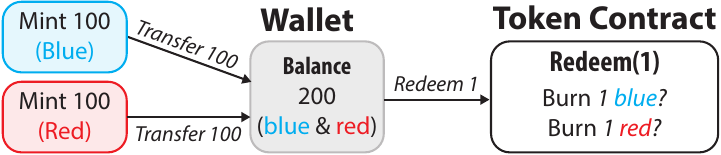}
    \caption{Existing fungible token contracts cannot fairly update attribution during redemption.}
    \label{fig:motivation}
    \vspace{-1em}
\end{figure}

We illustrate these three challenges in the example in Figure~\ref{fig:motivation}.
Two minters, blue and red, mint 100 tokens each, resulting in an attribution of 100 tokens for each color.
Both minters then send all 100 of their newly minted tokens to a single user's wallet, and the user redeems 1 of these 200 tokens at the token contract.
The token contract must reduce the attribution of minted tokens to reflect the reduction in the global supply of tokens, but the fungibility of the tokens makes it impossible to decide which colored attribution to slash.
\begin{figure}
    \centering
    \begin{subfigure}[b]{\columnwidth}
        \begin{algorithmic}[1]
            \Procedure{Mint}{$C, q$}
                \State $mint_C \gets mint_C + q$
            \EndProcedure
        \end{algorithmic}
        \vspace{0.1em}
        \begin{algorithmic}[1]
            \Procedure{Burn}{$C, q$}
                \State $mint_C \gets mint_C - q$
            \EndProcedure
        \end{algorithmic}
        \caption{$Mint$ and $Burn$ respectively increase or decrease $mint_C$.}
        \label{fig:mint-and-burn}
    \end{subfigure}\hfill
    \vspace{0.3em}
    \begin{subfigure}[b]{\columnwidth}
        \begin{algorithmic}[1]
            \Procedure{Wrap}{$C, b$}
                \State {$float_C \gets float_C + b$}
                \State \Return $b$
            \EndProcedure
        \end{algorithmic}
        \caption{$Wrap$ increments $float_C$ and returns the wrapped quantity.}
        \label{fig:wrap}
    \end{subfigure}\hfill
    \vspace{0.3em}
    \begin{subfigure}[b]{\columnwidth}
        \begin{algorithmic}[1]
        \Procedure{Unwrap}{$C, f$}
            \State $q \gets min(float_C, f)$ \label{cap-unwrap}
            \State $float_C \gets float_C - q$
            \State \Return $q$
        \EndProcedure
        \end{algorithmic}
        \caption{$Unwrap$ accepts a quantity of float tokens as input and decrements the relevant color's float balance.}
        \label{fig:unwrap}
    \end{subfigure}\hfill
    \vspace{0.3em}
    \begin{subfigure}[b]{\columnwidth}
        \begin{algorithmic}[1]
            \Procedure{Defloat}{$f$}
                \State $colors \gets \lbrack\rbrack$
                \While {$f < 0$} \label{start-float-burn}
                    \State $C_{rand} \gets \text{random color}$ \label{randomcolor}
                    \State $f \gets Unwrap(C_{rand}, f)$
                    \State $colors \gets colors + \lbrack C_{rand}q \rbrack$
                \EndWhile \label{end-float-burn}
                \State \Return $colors$
            \EndProcedure
        \end{algorithmic}
        \caption{$Defloat$ unwraps float tokens to be burned, returning the array of unwrapped tokens.}
        \label{fig:defloat}
    \end{subfigure}\hfill
    \vspace{0.3em}
    \begin{subfigure}[b]{\columnwidth}
        \begin{algorithmic}[1]
            \Procedure{Debit}{$q$}
                \If {$b_{local} + f_{local} \geq q$}
                \State $d_f \gets min(q, f_{local})$
                \State $b_{local} \gets b-(q - d_f)$
                \State $f_{local} \gets f - d_f$
                \State \Return $C_{local} (q - d_f) | d_f$
                \Else
                \State Revert
                \EndIf
            \EndProcedure
        \end{algorithmic}
        \caption{$Debit$ subtracts tokens from the sender's balance, exhausting float tokens before debiting colored tokens.}
        \label{fig:debit}
    \end{subfigure}\hfill
    \vspace{0.3em}
    \begin{subfigure}[b]{\columnwidth}
        \begin{algorithmic}[1]
        \Procedure{Credit}{$Cb|f$, $policy$}
            \If {$C_{local} = C$} 
            \State $b_{local} \gets b_{local} + b$ \label{credit-naive-start}
            \State $f_{local} \gets f_{local} + f$ \label{credit-naive-end}
            \ElsIf {$policy = \texttt{self}$}
            \State $f_{local} \gets f_{local} + Wrap(Cb|f)$ \label{credit-self}
            \ElsIf {$b_{local} \geq b$} \label{credit-floatminimized-start}
            \State $f_{local} \gets Wrap(C, b) + f + f_{local}$
            \Else
            \State $f_{local} \gets Wrap(C_{local}, b_{local}) + f + f_{local}$
            \State $C_{local} \gets C$
            \State $b_{local} \gets b$ \label{credit-floatminimized-end}
            \EndIf
        \EndProcedure
        \end{algorithmic}
        \caption{$Credit$ credits the receiver, wrapping tokens based on the specified policy.}
        \label{fig:credit}
    \end{subfigure}\hfill
    \label{fig:redeem}
    \caption{Token contract methods.}
\end{figure}
\section{Lossy token color encoding}
\label{sec:1-lossless-color-encoding}
We postulate that a practical solution to the fungible token coloring problem should have constant storage-complexity, and thus focus primarily on exploring constant space algorithms to track token coloring.
Any constant space-complexity solution to the fungible \Problem{} problem must be lossy, meaning information about all but a fixed number of colors will be lost during each token transfer.
We propose a class of constant space-complexity fungible token coloring algorithms, each notated as ColorFloat$_K$ where K is the number of colors that are encoded losslessly in each transaction.
In this section, we begin by presenting the algorithm for ColorFloat$_1$ (or just ``ColorFloat''), then generalize it to arbitrary $K$.
We argue that ColorFloat$_1$ is the most practical point in this tradeoff space.

\subsection{Token contract}
\label{sec:token-contract}
Minting and transferring of colored fungible tokens is controlled by the \emph{token contract}, which defines six functions: $Mint$, $Burn$, $Wrap$, $Unwrap$, $Transfer$, and $Cleanup$.
$Mint$ and $Burn$ (Figure~\ref{fig:mint-and-burn}) control $mint_C$ by incrementing or decrementing it respectively.

In addition to tracking $mint_C$, the token contract tracks what we term the \emph{float} of each color $float_C$.
To implement lossy color encoding, we introduce the concept of a \emph{float token}, which can conceptually be thought of as a colored token that has been wrapped into an \emph{uncolored} token.
This is the mechanism by which we implement lossy color encoding, converting a set of colored tokens into float tokens to be represented as a single quantity.
$Wrap$ and $Unwrap$ (Figures~\ref{fig:wrap} and \ref{fig:unwrap} respectively) control the wrapping and unwrapping of float tokens, incrementing and decrementing the float balance respectively of the specified color.
$Wrap$ is always invoked on a finite quantity of colored tokens, thus limiting the value of $float_C$ to be less than or equal to $mint_C$.
$Defloat$ (Figure~\ref{fig:defloat}) implements a mechanism to choose colors to unwrap.

Finally, the token contract implements $Debit$ and $Credit$ to facilitate token transfers (Figures~\ref{fig:debit} and \ref{fig:credit} respectively) between users.
Transfers introduce entropy into the system in the form of float tokens, while slashing balances that include float tokens reduces entropy in the system.
The key idea of ColorFloat is to penalize minters with high float balances (i.e., high entropy contribution) by increasing the expected value of tokens unwrapped to their color during redemption.
Minters can easily reduce their entropy by unwrapping float tokens they receive.

\subsection{Transfer algorithm}
\label{sec:algorithm}
We leverage the float token to decrease the space complexity of each token transfer from $O(N)$ to $O(1)$.
ColorFloat encoded balances are represented by a 3-tuple of \lbrack main color identifier ($C$), main color balance $b$, float balance $f$\rbrack, notated as $Cb|f$.
Each balance losslessly stores the balance of the main color, but lossily consolidates the balances of all other colors via the float token.
On every transfer, the recipient's recoloring policy decides a \emph{single} color to track losslessly, and wraps all other colored tokens into float tokens.

\begin{figure}
    \centering
    \begin{tabular}{c|r r}
        $Defloat(10)$ & Token & \multicolumn{1}{l}{Contract}\\
        \hline
        Initial state             & $mint_{C_1}: 10$ & $float_{C_1}: 4$\\ 
                                  & $\cdots$ & $\cdots$ \\
                                  & $mint_{C_2}: 10$ & $float_{C_2}: 8$\\
        \vspace{-0.95em} \\ \hline
        $C_1 \gets $ random color   & \\
        \hline
        $Unwrap(C_1, 10) \rightarrow 4$      & $mint_{C_1}: 10$ & $float_{C_1}: {\color{red} 0}$\\
                                  & $\cdots $\\
                                  & $mint_{C_2}: 10$ & $float_{C_2}: 8$\\
        \hline
        $C_2 \gets $ random color   & \\
        \hline
        $Unwrap(C_2, 6) \rightarrow 6$     & $mint_{C_1}: 10$ & $float_{C_1}: 0$\\
                                  & $\cdots $\\ 
                                  & $mint_{C_2}: 10$ & $float_{C_2}: {\color{red} 2}$\\
        \vspace{-0.95em} \\ \hline
        Colors:                                & $mint_{C_1}: 10$ & $float_{C_1}: 0$\\
        $\lbrack (C_1, 4), (C_2, 6) \rbrack $    & $\cdots$\\
                                               & $mint_{C_2}: 10$   & $float_{C_2}: 2$\\
    \end{tabular}
    \caption{$Defloat$ unwraps the float of randomly chosen colors to fulfill the request.}
    \label{fig:defloat-example}
    \vspace{-1em}
\end{figure}

\begin{figure*}
    \centering
    \begin{tabular}{c | l l l | l l}
         Action                             & Alice                 & Bob                   & Carol                 & Mint                      & Float                 \\
         \hline
         Alice mints 80 \textcolor{blue}{blue} tokens         & $\texttt{\textcolor{blue}{B}}80|0$     &                       &                       & $\texttt{\textcolor{blue}{B}}: 80$         & \\
         \hline
         Alice transfers 40 tokens to Bob   & $\texttt{\textcolor{blue}{B}}40|0$      & $\texttt{\textcolor{blue}{B}}40|0$      &                       & $\texttt{\textcolor{blue}{B}}: 80$         & \\
         \hline
         Carol mints 80 \textcolor{magenta}{pink} tokens         & $\texttt{\textcolor{blue}{B}}40|0$      & $\texttt{\textcolor{blue}{B}}40|0$      & $\texttt{\textcolor{magenta}{P}}80|0$     & $\texttt{\textcolor{blue}{B}}: 80$         & \\
                                            &                       &                       &                       & $\texttt{\textcolor{magenta}{P}}: 80$         & \\
         \hline
         Bob transfers 10 tokens to Carol   & $\texttt{\textcolor{blue}{B}}40|0$      & $\texttt{\textcolor{blue}{B}}30|0$      & $\texttt{\textcolor{magenta}{P}}80|10$    & $\texttt{\textcolor{blue}{B}}: 80$         & $\texttt{\textcolor{blue}{B}}: 10$      \\
         ($\texttt{\textcolor{blue}{B}}10$ converted to float)&                       &                       &                       & $\texttt{\textcolor{magenta}{P}}: 80$         & \\
         \hline
         Carol transfers 30 tokens to Alice & $\texttt{\textcolor{blue}{B}}40|30$     & $\texttt{\textcolor{blue}{B}}30|0$      & $\texttt{\textcolor{magenta}{P}}60|0$      & $\texttt{\textcolor{blue}{B}}: 80$           & $\texttt{\textcolor{blue}{B}}: 10$      \\
         ($\texttt{\textcolor{magenta}{P}}20|10$ debited, $\texttt{\textcolor{blue}{B}}0|30$ credited)&     &                       &                                        & $\texttt{\textcolor{magenta}{P}}: 80$           & $\texttt{\textcolor{magenta}{P}}: 20$      \\
         \hline
         Alice unwraps 20 float into \textcolor{blue}{blue}    & $\texttt{\textcolor{blue}{B}}60|10$     & $\texttt{\textcolor{blue}{B}}30|0$      & $\texttt{\textcolor{magenta}{P}}60|0$      & $\texttt{\textcolor{blue}{B}}: 80$         & $\texttt{\textcolor{blue}{B}}: \textcolor{red}{0}$       \\
         (10 \textcolor{blue}{blue} and 10 \textcolor{magenta}{pink} randomly chosen)&                       &                       &                       & $\texttt{\textcolor{magenta}{P}}: 80$         & $\texttt{\textcolor{magenta}{P}}: \textcolor{red}{10}$      \\
         \hline
         Bob burns 20 tokens (all \textcolor{blue}{blue})     & $\texttt{\textcolor{blue}{B}}60|10$     & $\texttt{\textcolor{blue}{B}}10|0$      & $\texttt{\textcolor{magenta}{P}}60|0$      & $\texttt{\textcolor{blue}{B}}: 60$          & $\texttt{\textcolor{blue}{B}}: 0$       \\
                                            &                       &                       &                       & $\texttt{\textcolor{magenta}{P}}: 80$         & $\texttt{\textcolor{magenta}{P}}: 10$      \\
         \hline
         Alice burns 10 tokens              & $\texttt{\textcolor{blue}{B}}50|0$      & $\texttt{\textcolor{blue}{B}}10|0$      & $\texttt{\textcolor{magenta}{P}}60|0$      & $\texttt{\textcolor{blue}{B}}: 60$          & $\texttt{\textcolor{blue}{B}}: 0$       \\
         (10 float automatically unwrapped to \textcolor{magenta}{pink}) &                       &                       &                       & $\texttt{\textcolor{magenta}{P}}: 70$          & $\texttt{\textcolor{magenta}{P}}: \textcolor{red}{0}$       \\
    \end{tabular}
    \caption{Three parties minting, exchanging, unwrapping, and burning tokens. Defloating is marked in \textcolor{red}{red}.}
    \label{fig:transfer-example}
    \vspace{-1em}
\end{figure*}

Transfers are mediated by the token contract through the $Transfer$ method.
For illustration purposes, we define the $Transfer$ method as the sequential invocation of the $Credit$ and $Debit$ subroutines in a single atomic transaction (Figures~\ref{fig:debit} and \ref{fig:credit}).
Each transfer involves two parties, a sender and receiver, with the sender transferring some quantity $q$ of tokens to the receiver.
$Debit$ then subtracts $q$ tokens from the sender's balance.
This operation first draws $f$ tokens from the float balance, then, if the float balance is exhausted, it consumes $b$ tokens from the main color balance for a final debited balance of $(c_sb|f)$.
$Credit$ runs after $Debit$, receiving as arguments the debited balance $c_sb|f$ and a policy for resolving color conflicts: \texttt{self} or \texttt{float-minimized}.
For brevity, we specify only these two policies in this paper, but any policy desired can be implemented by modifying the implementation of $Credit$.
If the main color of the debited balance matches the main color of the receiver, the balances are merged as-is (shown in Figure~\ref{fig:credit} lines~\ref{credit-naive-start}--\ref{credit-naive-end}).
If the main colors differ and the conflict resolution policy is set to \texttt{self}, $Credit$ wraps the entire debited balance to float before adding it to the receiver's balance (shown in line~\ref{credit-self}).
If the conflict resolution policy is set to \texttt{float-minimized}, the color with the larger balance is chosen to replace the receiver's color, with the smaller balance wrapped into float as illustrated in lines~\ref{credit-floatminimized-start}--\ref{credit-floatminimized-end}.

Crediting and debiting will never wrap more tokens of a color than the total mint of that color, which enforces the invariant that $\forall{C}, float_C \leq mint_C$.

When a token balance ($Cb|f$) is burned, the main color can be burned by calling $Burn(C, b)$.
However, the float tokens have no color and thus cannot be burned with the $Burn$ function.
$Defloat$ (Figure~\ref{fig:defloat}) facilitates burning of float tokens by unwrapping them into random colors.
The key observation that makes this scheme fair is that a high float balance for a given color implies a high rate of attrition for that token.
Therefore, minters should be incentivized to minimize this introduction of entropy by creating demand for users to configure that minter's color as their main color on the token contract.
As such, random selection of colors to unwrap float tokens into results in a proportional relationship between a minter's float balance and the expected value of tokens chosen by $Defloat$.

Figure~\ref{fig:defloat} lines \ref{start-float-burn}--\ref{end-float-burn} show this process when unwrapping $f$ float tokens.
$Defloat$ randomly selects a color, unwraps up to $f$ tokens into that color, and repeats until all $f$ tokens have been unwrapped.
After all $f$ float tokens are unwrapped into colored tokens, the main balance and the newly unwrapped tokens are burned.
The random color selection on line \ref{randomcolor} can be implemented onchain by using a pseudorandom hash of some user and/or transaction-specific metadata (e.g., using the hash of the user's public key and transaction parameters), which we illustrate in Figure~\ref{fig:defloat-example}.
After calculating which colors should be burned in which quantities, the tokens can be burned in the same transaction.
Note that $Unwrap$ will never unwrap more tokens of a color $C$ than $float_C$, as shown in Figure~\ref{fig:unwrap} line~\ref{cap-unwrap}.
An example of ColorFloat is shown in Figure~\ref{fig:transfer-example}, with three users exchanging and redeeming two colors of tokens (blue and pink).

\vspace{0.5em}
\noindent{\textbf{ColorFloat$_K$}} is a generalization of ColorFloat$_1$ that encodes balances as a (2k+1)-tuple encoded as $C_1b_1|C_2b_2|\dots|C_kb_k|f$.
In ColorFloat$_K$, debit must enable users to specify which colors to debit after the float is exhausted, and credit must implement a more expressive policy for resolving color conflicts.
For example, debit can accept a ranked list of colors to preferentially subtract from, and credit could losslessly encode the $K$ largest balances and wrap the remaining colors into float tokens.
The generalization of ColorFloat$_1$ to ColorFloat$_K$ depends largely on domain-specific implementation details (recoloring policy) so we omit a formal discussion of ColorFloat$_K$ for brevity.

\begin{figure}
    \centering
    \includegraphics[width=0.95\columnwidth]{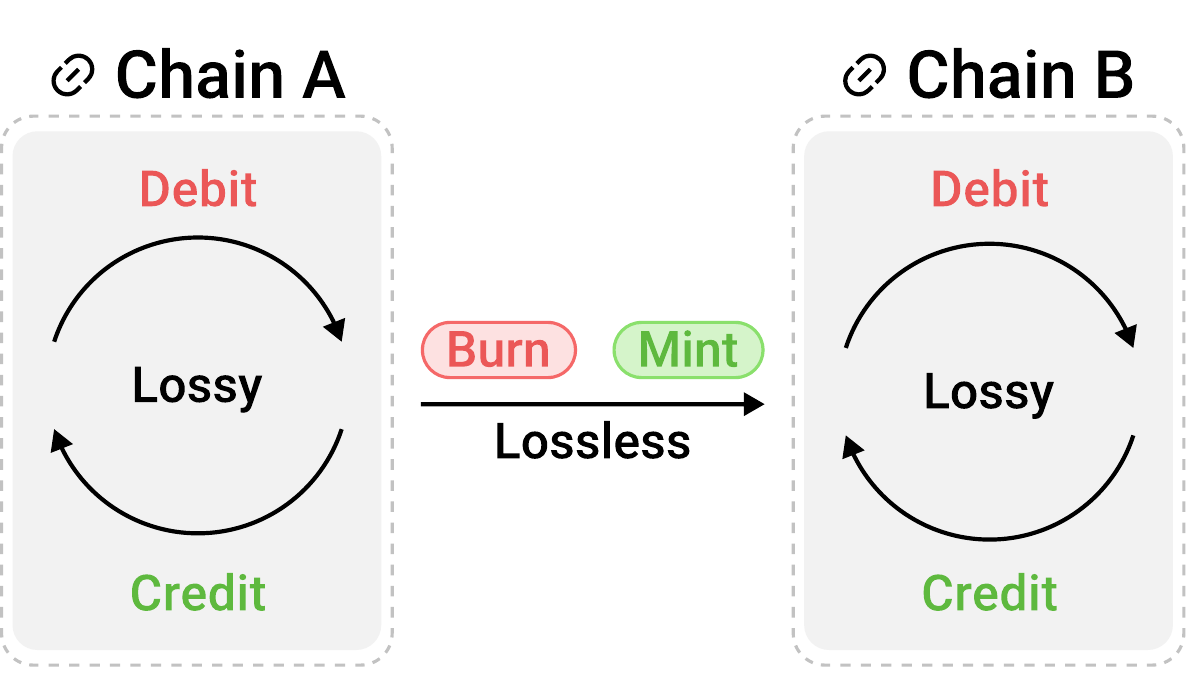}
    \caption{Transfers within a chain are lossy, but transfers between chains are lossless.}
    \label{fig:crosschain-transfer}
    \vspace{-1em}
\end{figure}
\vspace{0.5em}
\noindent{\textbf{Crosschain token transfers}} must be lossless, as cost constraints prevent float balances from being synchronized across multiple chains.
This, in turn, makes it impossible to fairly burn tokens that have been transferred to a different chain.
We suggest that crosschain token transfers losslessly encode a fixed number of colored tokens, which can be dynamically updated to reflect a set of minters who have the highest circulation across both chains in the transaction.
When tokens of a given color are transferred from a source chain to a destination chain.
However, there is a reduction in the mint on the source chain with a corresponding increase in the mint on the destination chain; however, the \emph{circulation} of the token does not change.
Conceptually, crosschain transfers are very simple: debit and burn a non-float balance on source, then mint and credit the corresponding balance on destination (illustrated in Figure~\ref{fig:crosschain-transfer}). 
To convert a balance with a nonzero float to a non-float balance, the user can \emph{Unwrap} any float tokens they hold to their wallet's main color.
Crosschain transfers can be facilitated by atomically moving tokens from the source chain to the destination chain with instant guaranteed finality using a messaging protocol such as LayerZero~\cite{layerzero-whitepaper}.
\section{Conclusion}
\label{sec:conclusion}
We presented the ColorFloat family of algorithms, solving the fungible token coloring problem by implementing lossy-- but fair-- attribution of tokens to minters.
The constant space complexity of ColorFloat enables the practical onchain application of these algorithms.
Multiple different-colored tokens are lossily aggregated into a single quantity by wrapping them into \emph{float} tokens, introducing entropy as less-desirable tokens are wrapped into float tokens.
These token colors that introduce entropy into the system are fairly penalized when tokens are burned, with colors that were wrapped in larger quantities more likely to have large portions of their mint slashed during token burning.
We believe this algorithm can be used in a variety of blockchain applications by allowing multiple minters to gain fair attribution while contributing value to the same token ecosystem.

{\footnotesize
\bibliographystyle{acm}
\bibliography{bib}}

\end{document}